\documentstyle[12pt]{article}          

\def	\eqnum		#1{(\ref{#1})}       
\def	\scite		#1{$^{\cite{#1}}$}     
	\newdimen\eqskip
	\newdimen\txtskip
	\baselineskip=\txtskip
	
\begin{document}

\begin{flushright}
Fermilab-Pub-94/273-T\\
hep-ph/9408397\\
August 30, 1994
\end{flushright}

\vspace*{1in}
\begin{center}
       	{ \Large  \bf \sc
STATUS OF THE SOLAR NEUTRINO PUZZLE
	}

\vskip 0.8in

{\it STEPHEN PARKE }\\[0.05in]
{\small
parke @ fnal.gov \\[0.1in]
Department of Theoretical Physics\\
Fermi National Accelerator Laboratory\\
P.O. Box 500, Batavia, IL 60510, U.S.A.} \\[0.2in]
\end{center}

\vskip 1.0in
\begin{abstract}
Using the latest results from the solar neutrino experiments and
a few standard assumptions,
I show that the popular solar models are ruled out at the
3$\sigma$ level
or at least {\em two} of the experiments are incorrect.
Alternatively, one of the assumptions could be in error.
These assumptions are spelled out in detail
as well as how each one affects the argument.
\end{abstract}

\def    \He          {\mbox{$^{4}He$}}
\def    \Be		{\mbox{$^{7}Be$}}
\def    \Bo		{\mbox{$^{8}B$}}
\def    \nupp		{\mbox{$\nu_{e}^{pp}$}}
\def    \nupep		{\mbox{$\nu_{e}^{pep}$}}
\def    \nuBe		{\mbox{$\nu_{e}^{^7 {Be}}$}}
\def    \nuBo		{\mbox{$\nu_{e}^{^8 B}$}}
\def    \f		{\mbox{$\phi$}}
\def    \fpp		{\mbox{$\f^{pp}$}}
\def    \fpep		{\mbox{$\f^{pep}$}}
\def    \fCNO		{\mbox{$\f^{CNO}$}}
\def    \fBe		{\mbox{$\f^{^{7}{Be}}$}}
\def    \fBo		{\mbox{$\f^{^{8}B}$}}
\def    \nfpp		{\mbox{${\Phi}^{pp}$}}
\def    \nfpep		{\mbox{${\Phi}^{pep}$}}
\def    \nfBe		{\mbox{${\Phi}^{^{7}{Be}}$}}
\def    \nfBo		{\mbox{${\Phi}^{^{8}B}$}}
\def    \SeC		{\mbox{$S^{ex}_{Cl}$}}
\def    \SeH		{\mbox{$S^{ex}_{H_2O}$}}
\def    \SeD		{\mbox{$S^{ex}_{Home}$}}
\def    \SeK		{\mbox{$S^{ex}_{Kam}$}}
\def    \StC		{\mbox{$S^{th}_{Cl}$}}
\def    \StH		{\mbox{$S^{th}_{H_2O}$}}
\def    \StG		{\mbox{$S^{th}_{Ga}$}}
\def    \SeG		{\mbox{$S^{ex}_{Ga}$}}
\def    \SeSe		{\mbox{$S^{ex}_{Sage}$}}
\def    \SeGx		{\mbox{$S^{ex}_{Gallex}$}}
\def    \X		{\mbox{$\chi^2$}}
\def    \BP		{\mbox{Bahcall \& Pinsonneault}}
\def    \TCL		{\mbox{Turck-Chi\`{e}ze \& Lopes}}

\def   	\bea            {\begin{eqnarray}}
\def   	\eea            {\end{eqnarray}}
\def   	\beq            {\begin{equation}}
\def   	\eeq            {\end{equation}}
\def 	\nn		{\nonumber}

\newpage
The solar neutrino puzzle has been with us for many years and an excellent
review
of the subject can be found in ref.\cite{Bbook}.
Recently the Gallium solar neutrino experiments have reduced their
uncertainties
so that a very simple argument can be made demonstrating the
difficulty of explaining
the experimental results using only known physics.
The simplest form of this argument is presented in this paper. Additional
features could be introduced which would strengthen this argument but
for the sake of simplicity and clarity they have not been included.

The argument, first used in ref.\cite{first},
makes the following assumptions about the sun,
neutrino properties and neutrino interaction cross sections:
\begin{itemize}
\item
The pp-solar-cycle is the dominant energy source of the Sun.
\item
The Sun is in a quasi-equilibrium, i.e. the solar luminosity
a few million years from now will be approximately the same as today.
\item
The neutrinos are unaffected during their propagation
from production in the solar core to their detection at the earth.
\item
The neutrino interaction cross sections for
the three types of experiments are correct.
\end{itemize}
With these four assumptions the main contributions to the solar
neutrino experiments are determined by
two parameters, the \Be~ and \Bo~ neutrino fluxes.
Therefore with three solar neutrino results one can compare the
standard solar models with the experimental results taken two at a time.

The main sequence of reactions that make up pp-solar-cycle
can be summarized as follows;
\bea
	\label{rpp}
		4p ~+~ 2e^- 	& \rightarrow & \He ~+~ 2 \nupp \\
	\label{rb7}
				& \rightarrow & \He ~+~ \nupp ~+~ \nuBe \\
	\label{rb8}
				& \rightarrow & \He ~+~ \nupp ~+~ \nuBo.
\eea
The total energy release in these reactions is 26.73 MeV but the
$\nupp, \nuBe,$ and $\nuBo$
carry off on average 0.265, 0.861 and 7 MeV respectively. Therefore
the energy release, not including the average neutrino energies, is
26.2, 25.6 and 19.5 MeV.

If the solar luminosity, $L_\odot$, is approximately constant over
a few million year time scale then there is a relation between
the current solar luminosity and the current solar neutrino fluxes, $\Phi^i$.
This relation can be written as
\bea
L_\odot & = & 13.1~(\nfpp - \nfBe - \nfBo)
  + 25.6~\nfBe + 19.5~\nfBo. \nn
\eea
For convenience it is useful to normalize the neutrino fluxes to those of
the solar model of Bahcall and Pinsonneault \scite{BP},
\beq
\f^i ~=~ \Phi^i / \Phi^i_{BP}
\eeq
where
\bea
\Phi^{pp}_{BP}  & = &  6.0 \times 10^{10} ~cm^{-2} sec^{-1} \nn \\
\Phi^{^7Be}_{BP} & = &  4.9 \times 10^{9} ~cm^{-2} sec^{-1} \nn \\
\Phi^{^8B}_{BP} & = & 5.7 \times 10^{6} ~cm^{-2} sec^{-1}. \nn
\eea
In these normalized flux units the solar luminosity constraint
is simply
\bea
	\label{lum}
	1 & = & 0.913~\fpp ~+~ 0.071~\fBe ~+~ 4 \times 10^{-5}~\fBo
\eea
This will be used to determine \fpp ~in terms of \fBe.

The contribution of the $\nupp, \nuBe$ and $\nuBo$
to the chlorine, water and gallium solar neutrino experiments is
\bea
	\label{tc}
	\StC & = & 6.2~\fBo ~+~ 1.2~\fBe  ~~~SNU  \\
	\label{th}
	\StH & = & \fBo  ~~~\Phi^{^8B}_{BP}      \\
	\label{tg0}
	\StG & = & 14~\fBo ~+~ 36~\fBe ~+~ 71~\fpp  ~~~SNU.
\eea
The coefficients in eq.\eqnum{tc}-\eqnum{tg0}
are determined using the assumptions that the state of the neutrinos
is unaffected by the passage from the solar core to the terrestrial detectors,
i.e. there is no change in the flavor, helicity or energy spectrum,
and that the neutrino interaction cross sections used are corrected.
The uncertainty on these cross sections is estimated to be a few per cent.

Using the luminosity constraint to eliminated the $\nupp~$ flux, the
contribution to the gallium experiments can be written as
\bea
	\label{tg}
	\StG & = & 14~\fBo ~+~ 30~\fBe ~+~ 78   ~~~SNU.
\eea
The additional contributions from other specifies of neutrinos is less than
10\%
in the standard solar models \scite{CNO}.

Over the past summer new results from the four solar neutrino
experiments have been reported.
The results for Homestake\scite{erc}, Kamiokande\scite{erh},
Gallex\scite{erg} and SAGE\scite{ers} are
\bea
	\SeD & = & 2.55 \pm 0.17 \pm 0.18  ~~~SNU   		\nn \\
	\SeK & = &  0.51 \pm 0.04 \pm 0.06 ~~~\Phi^{^8B}_{BP}   \nn  \\
	\SeGx & = & 79 \pm 10 \pm 6  ~~~SNU			\nn \\
	\SeSe & = & 69 \pm 11~^{+5}_{-7}  ~~~SNU		\nn
\eea
where the first uncertainty is statistical and second systematic.
To form a combined result for gallium, the mean and statistical errors
for SAGE and Gallex were combined in the standard way
but a common systematic error of 6 SNU was used.
Then the statistical and systematic errors
are combined in quadrature for each experimental result giving
\bea
	\label{ec}
       \SeC  & = & 2.55 \pm 0.25 ~~~SNU    \\
	\label{eh}
	\SeH & = & 0.51 \pm 0.072 ~~~\Phi^{^8B}_{BP}    \\
	\label{eg}
	\SeG   & = &   74 \pm 9.5  ~~~SNU.
\eea

These results are now used to fit the two parameters, \fBe~ and \fBo~,
of the model, eq.\eqnum{tc}, \eqnum{th} and \eqnum{tg}.
The \X~ variable was calculated for the four cases;
all three results together and the three ways of choosing two out of three.
Since the minimum value of \X~ occurs at negative values of \fBe~
for all four cases, the constraint
\bea
\fBe & \ge & 0
\eea
was imposed \scite{PDG}.
Table 1 contains the minimum value of \X~ with this constraint, which all
occur along $\fBe=0$, as well as the
value of \fBo~ at the minimum.

\begin{table}[b]
\begin{center}
\begin{tabular}{|c|c|c|}
\hline & & \\[-0.15in]
$L_\odot$ plus   &  $\chi^2_{min}$, $\fBe \ge 0 $ & $~~~~\fBo$  \\[0.05in]
\hline & & \\[-0.15in]
Cl~+~H$_2$O~+~Ga 	& 2.5	& 0.43 \\
Cl~+~H$_2$O    		& 1.4	& 0.44 \\
Cl~+~Ga 		& 1.0	& 0.41 \\
H$_2$O~+~Ga    		& 1.3	& 0.50 \\[0.05in]
\hline
\end{tabular}
\label{t1}
\caption[]{Minima of \X~ for $\fBe \geq 0$ and the value of \fBo ~at the
minimum.
All four minima occur along \fBe ~equals zero.}
\end{center}
\end{table}

Figures 1 through 4 are the contour plots of \X~ as function of
\fBe~ and \fBo~ for the four cases; Chlorine plus Water plus Gallium,
Chlorine plus Water, Chlorine plus Gallium and Water plus Gallium,
respectively.
The 1$\sigma$ to 5$\sigma$ contours are determined by
$\Delta \X=$ 2.3, 6.2, 11.8, 19.4 and 28.7, respectively \scite{gauss},
from the minimum
with $\fBe \ge 0$.
Also include on these plots are
the total theoretical ranges of the standard solar model predictions of
Bahcall \& Pinsonneault \scite{BP}, Turck-Chi\`{e}ze \& Lopes \scite{TCL}
and the ad hoc solar ``model'' where the central
temperature of the sun is a free parameter \scite{AH}.

Since the standard models of \BP~ and \TCL~
are consistent with our assumptions both of these models
are excluded by many sigma independent of which set of experimental results are
included.
Fig.2, using only Chlorine plus Water, is just a reformulation
of the argument by Bahcall and Bethe \scite{BB} but here the
exclusion is at the 5$\sigma$ confidence level.
Fig. 3 demonstrates a similar case for Chlorine Plus Gallium.
The least convincing case occurs with Water plus Gallium, Fig. 4,
and even then the two standard solar models are excluded
at almost the three sigma level. The ad hoc ``model,'' where the central
temperature of the sun is a free parameter, is excluded at the two sigma level
independent of which two experimental results are chosen.
It is worth noting that the case using Water plus Gallium excludes this model
at a higher level of confidence than either of the Chlorine plus Water
or the Chlorine plus Gallium cases.
Of course the case Chlorine plus Water plus Gallium gives the strongest
exclusion
to all models, Fig. 1.
If the contribution from the pep and CNO neutrinos
had been included the confidence level of all exclusions would have been
even stronger\scite{BeBo}.

The conclusion from these figures is that, given the assumptions delineated
above, either the standard solar models are
ruled out at the 3$\sigma$ level or at least two of the solar neutrino
experiments are incorrect\scite{R}.
Prior to the release of the latest experimental results,
only one of the solar neutrino experiments
needed to be incorrect to remove the discrepancy between the standard solar
models and the data.
Now, at least two experiments must be incorrect to
remove this discrepancy.
The probability that two independent experiments
are incorrect is considerably smaller than one.
This is a strong argument in favor of the conclusion
that one of the above assumptions is wrong
or that there is solar physics we do not understand.
One of the above assumptions is that neutrinos are
unaffected in their transition from
the solar interior to the terrestrial detectors.
The possibility that this assumption is incorrect has been discussed
by many authors who have suggested neutrino oscillations and/or neutrino
spin flip as explanations of the above discrepancy.

\begin{flushleft}{\bf Acknowledgements}\end{flushleft}
The author wishes to thank
C.~Albright, C.~Hill, T.~Kajita, C.~Quigg,
J.~Rosner and J.~Wilkerson
for discussions.
Fermilab is operated by the Universities Research Association
under contract with the United States Department of Energy.

\newpage
\begin{figure}[h]
\vspace{8cm}
\includegraphics{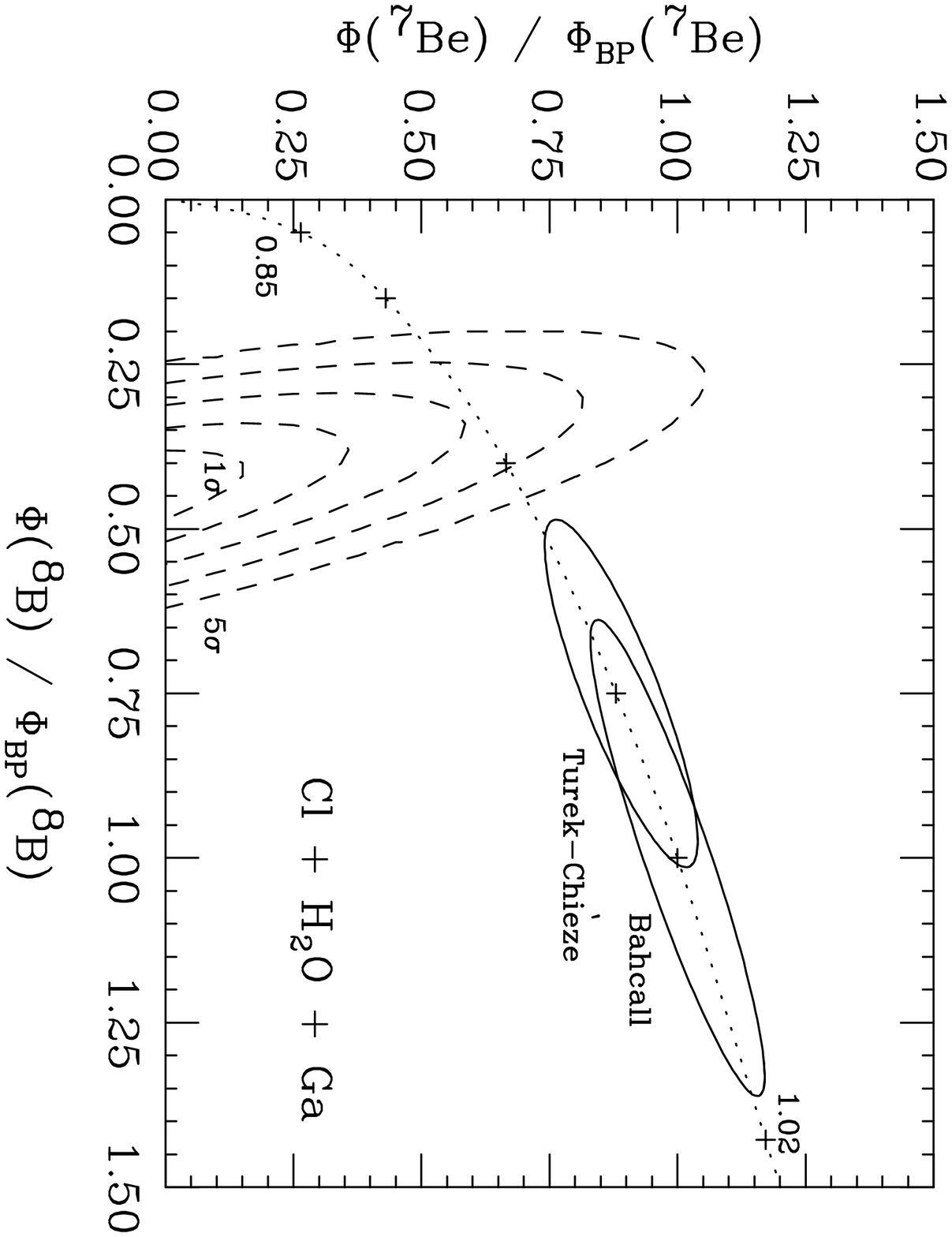}
\caption[]{
The \fBe~ verses \fBo~ plane using the results from the Chlorine, Water
and Gallium solar neutrino experiments.
The dashed curves are the 1$\sigma$ to
5$\sigma$ contours for the $\chi^2$ variable.
The solid ellipses are the predictions of the solar
models of Bahcall \& Pinsonneault and Turck-Chi\`{e}ze \& Lopes.
The dotted line is the curve $\fBe = (\fBo)^{8/18}$
and the crosses on this line corresponding to solar core temperature of
(0.85, 0.90, 0.95, 0.984, 1.00, 1.02) times the
core temperature of the Bahcall \& Pinsonneault's model.}
\label{chg0}
\end{figure}

\begin{figure}[h]
\vspace{8cm}
\includegraphics{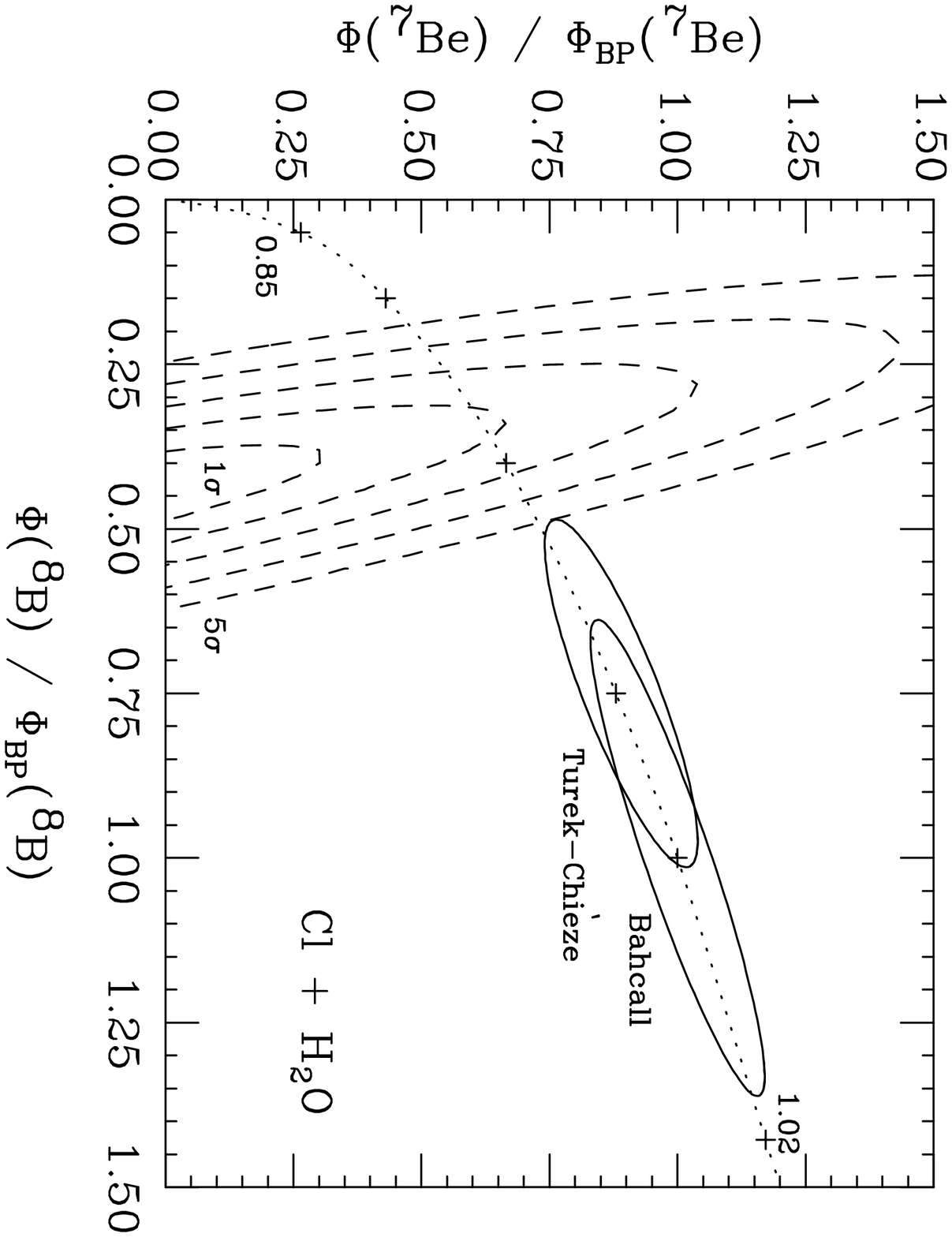}
\caption[]{
The \fBe~ verses \fBo~ plane using the results from the Chlorine and Water
solar neutrino experiments.
Curves as in Fig. 1.
}
\label{ch0}
\end{figure}

\newpage

\begin{figure}[h]
\vspace{8cm}
\includegraphics{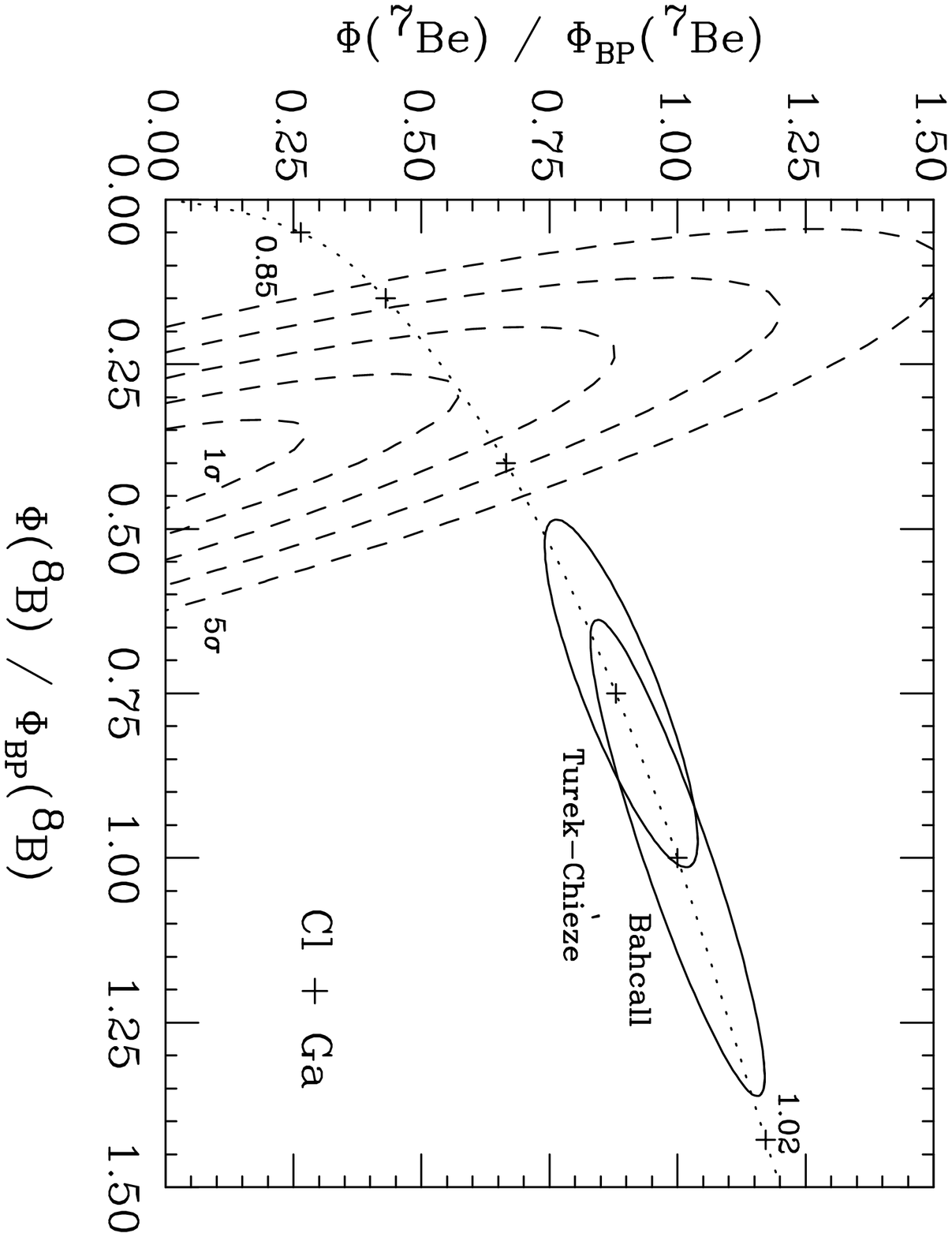}
\caption[]{
The \fBe~ verses \fBo~ plane using the results from the Chlorine
and Gallium solar neutrino experiments.
Curves as in Fig. 1.
}
\label{cg0}
\end{figure}

\begin{figure}[h]
\vspace{8cm}
\includegraphics{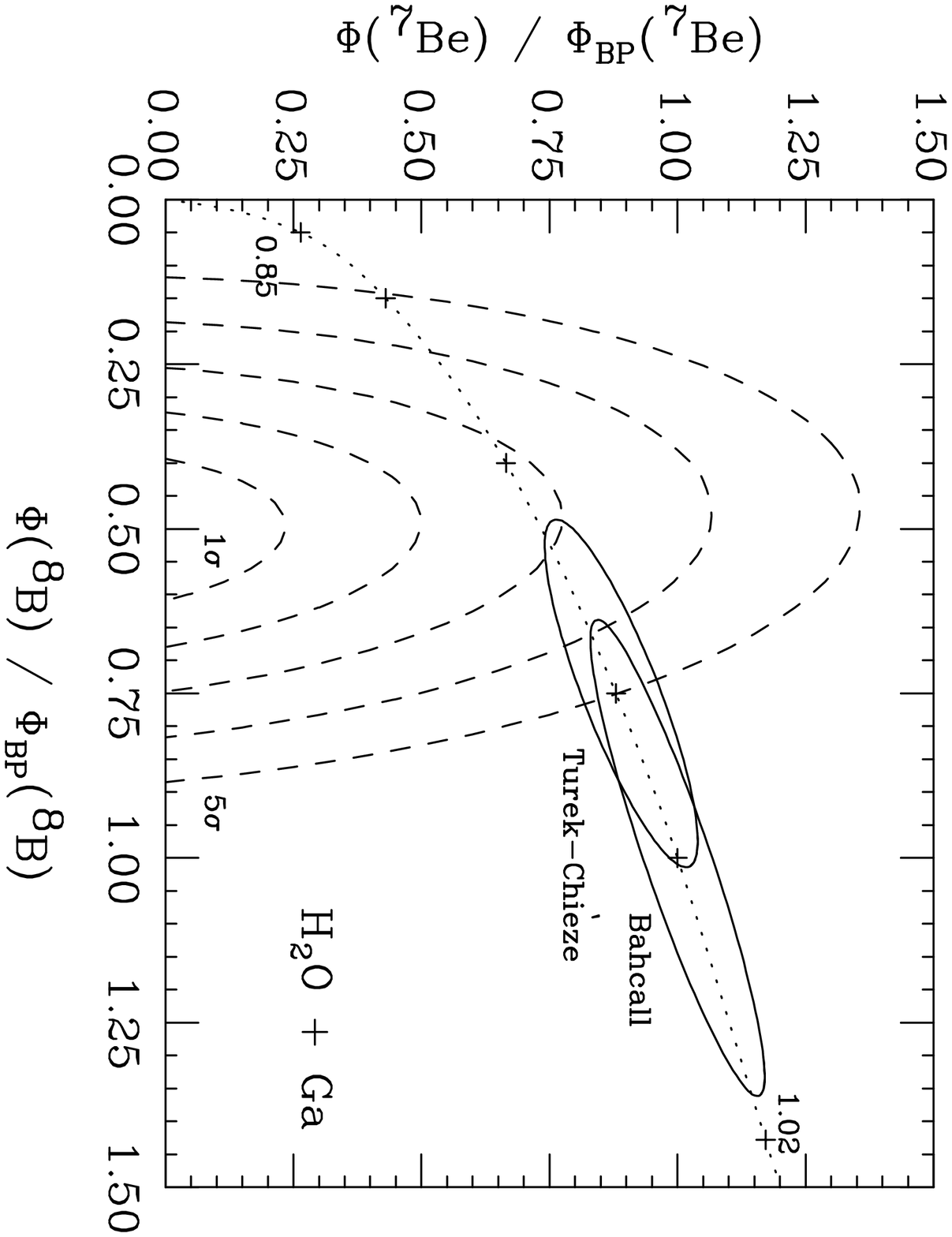}
\caption[]{
The \fBe~ verses \fBo~ plane using the results from the Water
and Gallium solar neutrino experiments.
Curves as in Fig. 1.
}
\label{hg0}
\end{figure}


\newpage
\begin{thebibliography}{99}
\parskip=0pt

\def    \astro     #1#2#3{{ Astrophys. J.} {\bf #1},  #3, (#2)}
\def    \nuke   #1#2#3{{ Nucl. Phys.} {\bf B#1},  #3, (#2)}
\def    \pl     #1#2#3{{ Phys. Lett.} {\bf #1B},  #3, (#2)}
\def    \prl    #1#2#3{{ Phys. Rev. Lett.} {\bf #1},  #3, (#2)}
\def    \pr     #1#2#3{{ Phys. Rev.} {\bf #1},  #3, (#2)}
\def    \prd    #1#2#3{{ Phys. Rev.} {\bf D#1},  #3, (#2)}
\def    \prep   #1#2#3{{ Phys. Rep.} {\bf #1},  #3, (#2)}
\def    \rmp    #1#2#3{{ Rev. Mod. Phys.} {\bf #1},  #3, (#2)}
\def    \zeit   #1#2#3{{ Z. Phys.} {\bf C#1},  #3, (#2)}
\def    \cmp    #1#2#3{{ Comm. Math. Phys.} {\bf #1},  #3, (#2)}
\def    \ibid   #1#2#3{{\it ibid.} {\bf #1}, #3, (#2)}
\def    \jetp   #1#2#3{{ JETP Lett.} {\bf #1},  #3, (#2)}
\def    \sovnuke #1#2#3{{ Sov. J. Nucl. Phys.} {\bf #1},  #3, (#2)}

\bibitem{Bbook}
J.~N.~Bahcall, Neutrino Astrophysics, Cambridge University Press.
\bibitem{first}
M.~Spiro and D.~Vignaud, \pl{242}{1990}{279},\\
V.~Castellani, S.~Degl'Immocenti and G.~Florentini,
Astron. Astrophys. {\bf 271}, 601, (1993),\\
N.~Hata, S.~Bludman and P.~Langacker. \prd{49}{1994}{3622}.
\bibitem{BP}
J.~N.~Bahcall and M.~H.~Pinsonneault, \rmp{64}{1992}{885}.
\bibitem{CNO}
Even if the CNO and pep neutrinos where included in the luminosity
constraint, the coefficients of \fCNO~ and \fpep~ would be positive
in eq. \eqnum{tg} so that any contribution from these sources would
strengthen the argument.
\bibitem{erc}
K.~Lande for the Homestake Collaboration, Neutrino 94, Israel, June 1994.
\bibitem{erh}
Y.~Suzuki for the Kamiokande III Collaboration, Neutrino 94.
\bibitem{erg}
T.~Kirsten for the Gallex Collaboration, Neutrino 94.
\bibitem{ers}
J.~Nico for the SAGE Collaboration, 27th ICHEP, United Kingdom, July 1994.
\bibitem{PDG}
This is the procedure recommended by the Particle Data Group,
\prd{45}{1992}{Part II} and gives a more conservative result
than not imposing this constraint.
\bibitem{gauss}
This assumes all errors are gaussian distributed!
\bibitem{TCL}
S.~Turck-Chi\`{e}ze and I.~Lopes, \astro{408}{1993}{347}.
\bibitem{AH}
In the ad hoc ``model,'' $\nfBe \sim T_c^{8}$
and $\nfBe \sim T_c^{18}$ where $T_c$ is the solar core temperature.
Therefore this model is represented by the line $\fBe = (\fBo)^{8/18}$.
\bibitem{BB}
J.~N.~Bahcall and H.~Bethe, \prl{65}{1990}{2233}.
\bibitem{BeBo}
Unfortunately this argument in this simple form is not strong
enough to exclude $\fBe \geq \fBo$ by more than about 1.2$\sigma$
unless all three results are include. Additional complications are needed.
\bibitem{R}
Similar conclusions, using a different argument, were presented in\\
J.~Bahcall, IASSNS-AST-94/37 preprint, and by\\ S~P.~Rosen
at the 27th ICHEP Conference, United Kingdom, July 1994.

\end{thebibliography}
\end{document}